\documentstyle [12pt,dina4p,epsfig]{article}

\begin{document}
\begin{titlepage}
\begin{flushleft}

\noindent
{\tt DESY 96-236    \hfill    ISSN 0418-9833} \\
{\tt November 1996} \\
\end{flushleft}
\vspace*{3.cm}
\begin{center}
\begin{LARGE}
{\bf
Determination of the Longitudinal Proton Structure Function 
{\boldmath $F_L(x,Q^2)$} at Low {\boldmath $x$} 
 \\ }

\vspace*{2.cm}
  H1 Collaboration \\
\end{LARGE}
\vspace*{4.cm}
{\bf Abstract:}
\begin{quotation}
\noindent
    A  measurement of the inclusive 
   cross section for the deep-inelastic scattering of positrons
   off protons at HERA is presented
   at momentum transfers $8.5 \leq Q^2 \leq 35~$GeV$^2$ and
   large inelasticity $y = 0.7$, i.e.
   for the Bjorken-$x$ range $0.00013 \leq x \leq  0.00055$.
   Using a next-to-leading order QCD fit to the structure
   function $F_2$ at lower $y$ values,  the 
   contribution of $F_2$ to  the measured 
   cross section at high $y$ is calculated and,
   by subtraction, the longitudinal 
   structure function $F_{L}$ is determined for the first time with
   an average value of $F_L=0.52 
   \pm 0.03$~(stat)$^{~+0.25}_{~-0.22}$ (syst)
   at $Q^2=15.4$~GeV$^2$ and $x=0.000243$.
\end{quotation}
\vfill
 submitted to Phys. Lett. {\bf B}
\cleardoublepage
\end{center}
\end{titlepage}
 \begin{flushleft}
 C.~Adloff$^{35}$,                
 S.~Aid$^{13}$,                   
 M.~Anderson$^{23}$,              
 V.~Andreev$^{26}$,               
 B.~Andrieu$^{29}$,               
 C.~Arndt$^{11}$,                 
 A.~Babaev$^{25}$,                
 J.~B\"ahr$^{36}$,                
 J.~B\'an$^{18}$,                 
 Y.~Ban$^{28}$
% ...
 \\
 \bigskip\bigskip
 {\it
% ...
$ ^2$ III. Physikalisches Institut der RWTH, Aachen, Germany$^ a$ \\
% ...
$ ^{11}$ DESY, Hamburg, Germany$^a$ \\
% ...
$ ^{13}$ II. Institut f\"ur Experimentalphysik, Universit\"at Hamburg,
                                     Hamburg, Germany$^ a$  \\
% ...
$ ^{18}$ Institute of Experimental Physics, Slovak Academy of
                Sciences, Ko\v{s}ice, Slovak Republic$^{f,j}$\\
% ...
$ ^{22}$ Physics Department, University of Lund,
                                               Lund, Sweden$^ g$ \\
 $ ^{23}$ Physics Department, University of Manchester,
                                          Manchester, UK$^ b$\\
% ...
$ ^{25}$ Institute for Theoretical and Experimental Physics,
                                                 Moscow, Russia \\
 $ ^{26}$ Lebedev Physical Institute, Moscow, Russia$^ f$ \\
% ...
$ ^{28}$ LAL, Universit\'{e} de Paris-Sud, IN2P3-CNRS,
                            Orsay, France\\
 $ ^{29}$ LPNHE, Ecole Polytechnique, IN2P3-CNRS,
                             Palaiseau, France \\
% ...
$ ^{35}$ Fachbereich Physik, Bergische Universit\"at Gesamthochschule
               Wuppertal, Wuppertal, Germany$^ a$ \\
 $ ^{36}$ DESY, Institut f\"ur Hochenergiephysik,
                              Zeuthen, Germany$^ a$\\
% ...
$ ^a$ Supported by the Bundesministerium f\"ur Bildung, Wissenschaft,
        Forschung und Technologie, FRG,
        under contract numbers 6AC17P, 6AC47P, 6DO57I, 6HH17P, 6HH27I,
        6HD17I, 6HD27I, 6KI17P, 6MP17I, and 6WT87P \\
 $ ^b$ Supported by the UK Particle Physics and Astronomy Research
       Council, and formerly by the UK Science and Engineering Research
       Council \\
% ...
$ ^f$ Supported by the Deutsche Forschungsgemeinschaft \\
 $ ^g$ Supported by the Swedish Natural Science Research Council \\
% ...
$ ^j$ Supported by VEGA SR grant no. 2/1325/96 \\
    } 
 \end{flushleft}

\newpage
\section{Introduction}
Precise measurements of the inclusive scattering cross section at 
the $ep$ collider HERA are important for the understanding of proton 
substructure. In the one-photon exchange approximation, which is valid
in the kinematic domain explored here, the deep 
inelastic  scattering (DIS)  cross section is 
given by the expression
\begin{equation}
        \frac{d^2\sigma}{dxdQ^2} = \frac{2\pi \alpha^2}{Q^4 x}
        \cdot [(2(1-y)+y^2) F_2(x,Q^2) -y^2 F_L(x,Q^2)].
        \label{sig}
  \end{equation}  
Here $Q^2$ is the squared four-momentum transfer, $x$ denotes the Bjorken 
scaling variable, $y=Q^2/sx$ is the inelasticity, with
$s$ the  $ep$ center of mass energy squared,
and $\alpha $ is the fine structure constant. The  structure 
functions $F_2$ and $F_L$ are related to the cross sections $\sigma_T$ and 
$\sigma_L$ for the interaction of transversely and longitudinally polarized 
virtual photons with protons. In the Quark Parton Model $F_2$ is the sum 
of quark and antiquark distributions multiplied by $x$ and
weighted with the square of the electric charges of the quarks, while $F_L$
 is predicted to be zero for spin 
$1/2$ partons \cite{CG}. In Quantum Chromodynamics (QCD) $F_L$ acquires a 
non zero value due to gluon radiation which is proportional to the strong 
coupling constant $\alpha_s$ \cite{AM} with possibly sizeable higher 
order corrections in QCD perturbation theory \cite{vN}. Measurements of 
$F_L$, expressed as the structure function ratio 
\begin{equation}
        R=\frac{F_L}{F_2-F_L} = \frac{\sigma_L}{\sigma_T}
        \label{R}
    \end{equation}
have been made by various fixed target lepton-hadron scattering 
experiments at higher $x$ values \cite{Bo, Mil}. This paper 
presents the first determination of $F_L(x,Q^2)$ at HERA in the deep 
inelastic region of $8.5 \leq Q^2 \leq 35~$GeV$^2$ and very small $x$ 
values between $1.3 \cdot 10^{-4}$ and $5.5 \cdot 10^{-4}$.

The H1 collaboration has recently reported a measurement of the  
structure function $F_2$ \cite{H1F} in the range $3 \cdot 10^{-5} \leq
x \leq 0.32$ and $1.5 \leq Q^2 \leq 5000 $~GeV$^2$,
using data taken in the year 1994. The measurement was 
restricted to $y$ values between 0.01 and 0.6 where the contribution of 
$F_2$ to the cross section, Eq. (\ref{sig}), dominates.
The $F_2$ values were extracted from the measured cross 
sections assuming theoretically computed values of $F_L$.
A  next-to-leading order (NLO) QCD analysis showed that
 the $F_2$ structure function can be well described by 
the DGLAP evolution equations \cite{DGLAP} 
in the kinematic range of the measurement.  

At high $y$ the  factors $Y_+ = 2(1-y)+y^2$ and $y^2$
multiplying $F_2$ and $F_L$, respectively,
are of comparable size. Therefore, the usual technique of extracting 
$F_2$ assuming a calculated $F_L$ is  reversed  and $F_L$  is determined
by subtraction of the $F_2$ contribution from the measured cross section.
The following procedure is applied. Our measurement of $F_{2}$  
\cite{H1F}, for $y < 0.35$, and fixed target data at
 larger $x$ \cite{BCDMS} are used to
extract the parton distribution functions which are evolved in 
$Q^{2}$ according to the NLO DGLAP evolution equations. This provides
predictions for the structure function $F_{2}$ in the high $y$ region 
which allow, by subtraction of the contribution of $F_{2}$ 
to the DIS cross section (cf. 
Eq. (1)), the determination of the longitudinal structure function
$F_{L}$ to be made.
  Note that the measurements of $F_{2}$ are well described by NLO QCD 
 over four orders of magnitude in $x$ and $Q^{2}$ while the evolution 
 required here extends the maximum $Q^{2}$ at fixed $x$  by a 
 factor of two only.
Nevertheless, since
an extended kinematic region is accessed here, where new effects could be
important, it can not be excluded that 
the structure function $F_2$ behaves differently than assumed.

Instead of subtracting the contribution of $F_2$ from the cross
section one could perform a cross section analysis using the 
QCD predictions for both structure functions $F_2$ and $F_L$.
This would be conceptually different to the method employed in this
paper because then an assumption would be required not only for
$F_2$ but also for $F_L$ which is less well known than $F_2$.

 A salient feature of 
the subtraction method is a partial cancelation of systematic errors 
because the cross sections at low and at high $y$ are measured
using one common set of data.  The experimental challenge is to 
measure the cross section at high $y$
 where the energy of the scattered positron $E_e'$
is comparatively low. The present measurement is made for 
$11 \geq E_e' \geq 6.5$~GeV, or $0.6 < y < 0 .78$, which is an extension of the
kinematic range covered by our previously published measurement
of $F_2$ \cite{H1F}.
An understanding of the trigger efficiency,
 positron identification, photoproduction
background and radiative corrections now becomes more demanding.

The paper is organized as follows. Section 2 discusses
 the cross section 
measurement with particular emphasis on the high $y$ region. 
 Section 3 
describes the QCD fit used to 
define the $F_2$ contribution for  subtraction and presents 
the final results. A short summary is given in
Section~4.
\section{Cross Section Measurement}
\subsection{Kinematics}
In 1994  HERA was operated with positrons  
of energy $E_e=27.5~$GeV   and protons of energy $E_p=820~$GeV.  
The event kinematics were reconstructed using the energy
of the  scattered positron 
$E_e'$ and the polar angle $\theta_e$ according to the relations
\begin{equation}
Q^2_e= \frac{E^{'2}_e \sin^2{\theta_e}}{ 1-y_{e}}
 \hspace*{2cm}
y_e=1-\frac{E_e'}{E_e}~\sin^2(\theta_e/2).
\label{qy}
\end{equation}
Here $\theta_e$ is defined with respect to the proton beam direction, 
defining the $z$ axis,   and
$x$ is calculated as $Q^2/sy$ with  $s=4E_eE_p$.  
At high $y$ the determination of $Q^2$ and $y$ from the reconstructed 
positron, rather than from the final state hadrons or a combination of 
both, is preferred because of the superior resolutions in 
$Q^2$ and $x$. The determination of the inclusive event kinematics
using the variables $E_e'$ and $\theta_e$ is subsequently referred to
as the ``electron method''.

The previously published analysis \cite{H1F} used the ``sigma 
method'' to determine $F_2(x,Q^2)$ for $y<0.15$. This method combines
the positron with the hadronic measurement by defining 
\begin{equation}
   Q^2_{\Sigma} = \frac{E^{'2}_e \sin^2{\theta_e}}{ 1-y_{\Sigma}}
   \hspace*{2cm}
   y_{\Sigma} = \frac{y_h}{ 1+y_h-y_e},
\end{equation}
which avoids the resolution degradation of $y_e$ at low
$y$. Here
\begin{equation}
       y_h=\frac{\sum_{i}(E_i-p_{z,i})}{2E_e},
       \label{yh}
\end{equation}
where $E_i$ and $p_{z,i}$ are the energy and longitudinal 
momentum component of a particle $i$. The summation extends over all 
hadronic final state particles and the masses are neglected. 

  The kinematic region of the $F_L$ measurement was
  limited by the constraints $0.6<y<0.78$, $155^o < \theta_e< 171^o$.  
  It was divided into six intervals of $Q^2$ with the limits 
  (7.5, 10.0, 13.3, 17.8, 23.7, 31.6, 42.2)~GeV$^2$ and
  with  central values
  chosen to be (8.5, 12.0, 15.0, 20.0, 25.0, 35.0)~GeV$^2$  or 
  the corresponding values of 
  $x=Q^2/sy$ at $y=0.7$.
\subsection{The H1 Detector}
The H1 detector~\cite{H1} is a nearly hermetic apparatus built to
investigate   high-energy $ep$ interactions at HERA. The
measurement of the inclusive deep inelastic cross section relies
essentially on the inner tracking chamber system and on the backward
electromagnetic and liquid argon calorimeters. A superconducting
 solenoid surrounds both the tracking system and the liquid argon
calorimeter, providing a uniform magnetic field of 1.15~T.

The energy of the  scattered positron  was measured in the backward 
electromagnetic calori\-meter (BEMC) behind which  a 
scintillator hodoscope (TOF) was placed to veto proton beam induced 
background interactions.  The identification of the scattered positron and the 
 measurement of the polar angle  made use of the  backward 
multiwire proportional 
chamber (BPC) which was attached to the BEMC. In the kinematic range of 
this $F_L$ measurement the positron angle was
 limited to $155^o<\theta_e<171^o$.
 For these angles the scattered 
positron traversed the inner cylindrical 
proportional chamber (CIP) which could therefore be included in the 
positron identification requirement.
The inner and outer proportional 
chambers (CIP at 18 cm radius and COP at 47 cm radius) were used in 
the trigger to reconstruct tracks of particles originating from the 
interaction region, and thus to reduce beam induced background events. The 
interaction vertex was determined with the central drift chambers and the 
hadronic final state was reconstructed with the Liquid Argon  
calorimeter and the tracking detectors. 

The luminosity was determined from the cross section of the 
elastic bremsstrahlung process, $ep \rightarrow ep\gamma$, 
measured with a precision of 1.5\%.
The integrated luminosity for this analysis is 1.25~pb$^{-1}$.
 The final state positron and the photon scattered at very low $Q^2$ 
can be detected in calorimeters (``electron and photon taggers'')
which are situated 
33~m  and  103~m from the interaction point in  the 
positron beam direction.

The use of the H1 detector for the inclusive DIS cross section 
measurement is further discussed in \cite{H1F}. A detailed technical
description of the apparatus can be found in \cite{H1}.
 
\subsection{Trigger}
The DIS event selection was based on  events triggered in the BEMC by 
an energy deposition of more than 6~GeV, combined with a TOF 
requirement and a valid CIP-COP track signal. The efficiency of the 
BEMC energy requirement was monitored using a central track trigger 
and found to be better than 99\% for the whole analysis region. The 
CIP-COP trigger required at least one track pointing to the  
interaction region.
The efficiency  was 96\% 
after all selection cuts. It was found to vary little 
over the region of acceptance  and to be well reproduced by the 
simulation of the trigger response. It was monitored
down to a positron energy of 7.5~GeV by an independent 
BEMC trigger and was evaluated  between 6.5 and 7.5~GeV 
by studying its dependence on $\theta_e$, on the hadronic angle 
 and on the charged track multiplicity
comparing simulation with data.

\subsection{Event Selection}
The event selection criteria are summarized in Table \ref{CUTS}. The positron 
 was identified as the most energetic cluster in the BEMC  associated 
 with
\begin{table}[tb] \centering 
\begin{tabular}{|c|}
\hline
6.5~GeV$ < E_e' < 11$~GeV  \\
$\epsilon_1 < 4~$cm  \\
$\epsilon_2 < 3.5~$cm  \\
$ \epsilon_3 < 5~$cm \\
$-25~$cm$ < z_{vtx} <35~$cm  \\
$ N_{tr} > 1 $  \\
\hline
\end{tabular}
\caption{\label{CUTS}
\sl Summary of event selection criteria.
For positron identification three estimators were used -
    $\epsilon_1$: reconstructed positron cluster radius in the BEMC;
    $\epsilon_2$: distance from the closest BPC hit to the     
         centroid of the positron cluster;
     $\epsilon_3$:  distance from the  positron candidate trajectory
      to the closest  active
     CIP pad (not used for tracks outside the CIP acceptance region).
   $z_{vtx}$ denotes the $z$ position of the reconstructed interaction
   vertex and $N_{tr}$ is the number of charged tracks reconstructed
   in the central drift chambers.}      
\end{table}
a signal in the preceding BPC and, if geometrically accessible, in the
CIP. For the determination of the event kinematics and background
suppression a vertex had to be reconstructed with
more than one track in the central drift chamber.

Deep inelastic events were generated using the DJANGO \cite{django}
program which is based on HERACLES \cite{HERAKLES} for the electroweak
interaction and on LEPTO \cite{LEPTO} to simulate the hadronic final
state. Photoproduction background was generated with the PHOJET
\cite{PHOJET} program. The detector 
response was simulated using a program based on GEANT \cite{GEANT}.
The simulated  events were subjected to the same
reconstruction and analysis chain as the  data. For comparisons with
experimental distributions, all simulated spectra were normalized
to the measured luminosity. 

Fig. 1a shows the distribution of the energy of the scattered positron
for the events passing all selection criteria in the high $y$
region. The experimental distribution is very well described by the
superposition of the simulated spectra from DIS events and of the 
photoproduction background, discussed below.

\begin{figure}[t]\centering
\begin{picture}(200,160) 
\put(-170,-150){\epsfig{file=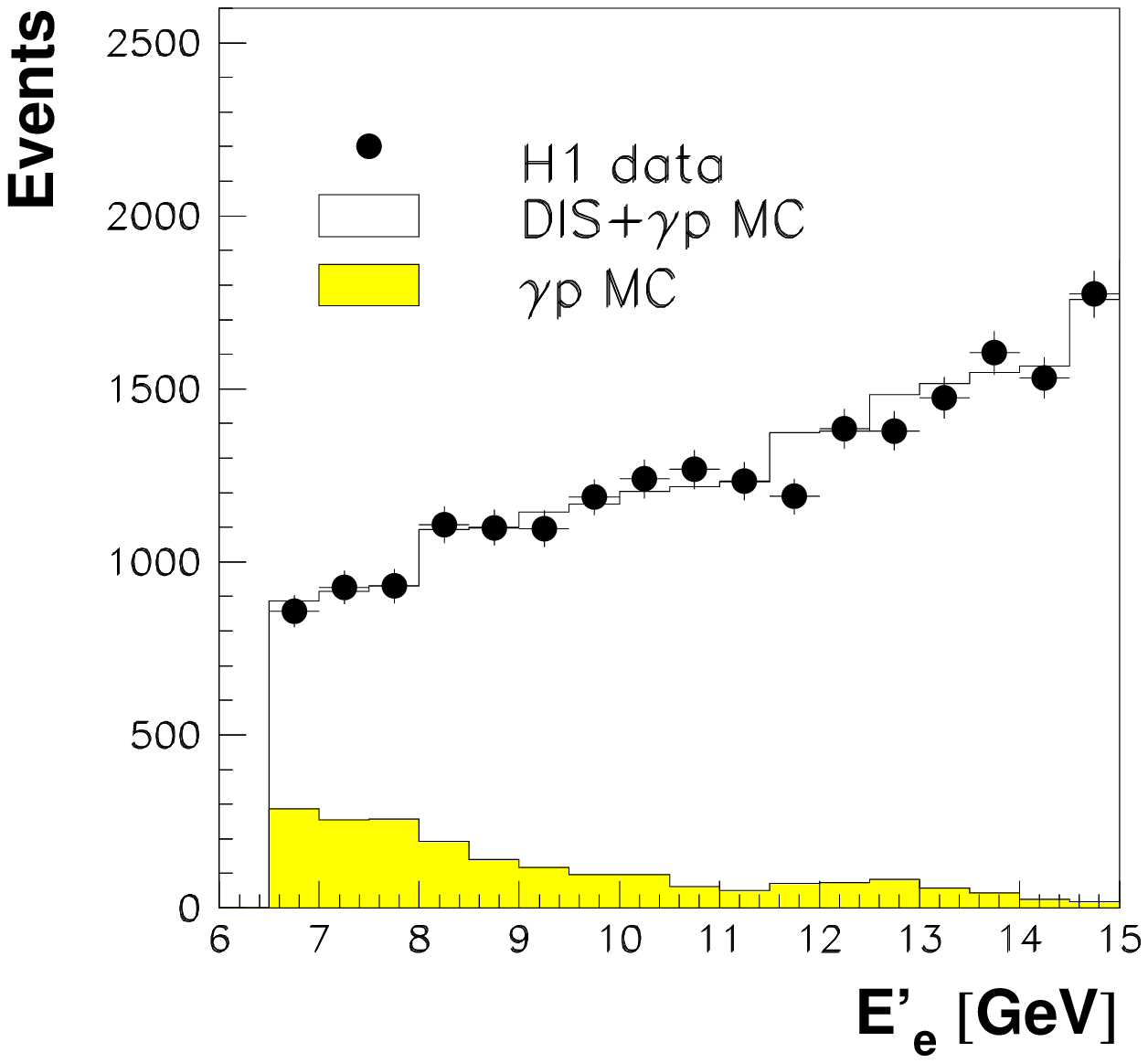,
width=10cm,height=15cm,bbllx=0pt,bblly=0pt,bburx=557pt,bbury=792pt}}
\put(40,30){a)}
\put(60,-150){\epsfig{file=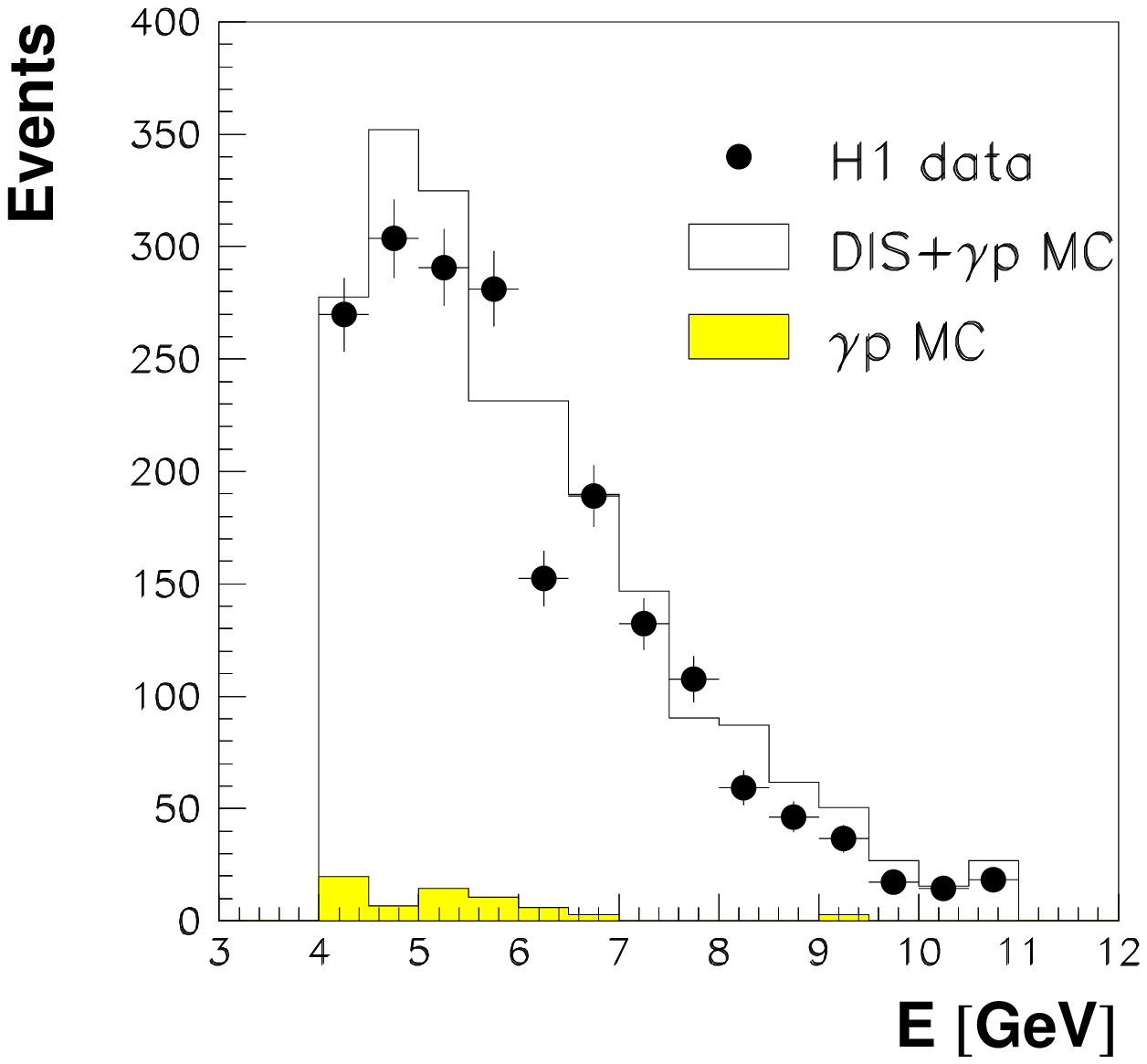,
width=10cm,height=15cm,bbllx=0pt,bblly=0pt,bburx=557pt,bbury=792pt}}
\put(250,30){b)}
\end{picture} 
\vspace*{2cm}
        \caption {\sl{Energy distributions of a) the highest energy
            and b)
       the next highest energy BEMC clusters  for the final data sample.
       The simulated spectra are normalized  to the 
       luminosity of the data.}}
        \protect\label{be}
\end{figure}

 For the genuine DIS events at high $y$, 
the current jet particles are on average emitted
backwards with respect to the proton beam direction.
Thus there is a possibility that the largest energy cluster is not due to 
the scattered positron but  to a hadronic 
energy deposition in the BEMC. In a Monte Carlo simulation  of 
DIS events 3\% of the 
selected positron candidates were found to be 
produced by the hadronic final state.
In more than 99\% of the simulated events the genuine positron
 was either the highest energy or second highest energy cluster. 
For a comparative study of 
the effect of misidentification at low energies, 
a positron finding algorithm was used which accepted an event even 
if the highest energy cluster failed to satisfy the selection 
conditions but 
the second highest energy cluster fulfilled them. The resulting cross section 
agreed to within 1\% with that based on the standard 
positron finding algorithm which used the highest energy cluster.
 Fig. \ref{be}b shows the BEMC
energy distribution of the cluster with second highest energy which is
well reproduced by the simulation.
The background due to photoproduction is small  
because there is only a small probability to generate 
two high energy clusters in the BEMC in such events.
\subsection{Photoproduction Background}
At low  energies and  for large polar angles
of the scattered positron
there are two major
sources of background in the candidate DIS events. Non-$ep$
background occurs due to  beam interactions with residual gas and 
beam-line elements. An effective filter against such
events is the requirement of a reconstructed event vertex in the
interaction region. The number of
remaining beam-induced background events was estimated  to be $2.5\%$
in the lowest $Q^2$ interval and below $1\%$ everywhere
else using non-colliding bunch events.

The second, more difficult, background source is  photoproduction,
 including low $Q^{2}$ DIS events, in which
the scattered positron escapes undetected along the beam pipe and
in which an energy cluster from the final state particles fakes a
positron  signal in the BEMC.
The typical characteristic of the $\gamma p$ background is a rapid
rise of the cross section towards lower cluster
energy.  Most frequently, the energy cluster  in the BEMC
is  produced by a $\pi^0$ decay to two photons or by charged
hadrons, mainly  $\pi^{\pm}$. 
Energy clusters due to neutral particles
are effectively removed by demanding a track pattern in the CIP and 
the requirement that a BPC hit coincide spatially with a BEMC cluster.
Hadronic clusters typically have large cluster
radii in the BEMC, and are rejected by the  cut on 
$\epsilon_1$, see Table 1.

\begin{figure}[hb]\centering
\begin{picture}(200,150) 
\put(-170,-220){\epsfig{file=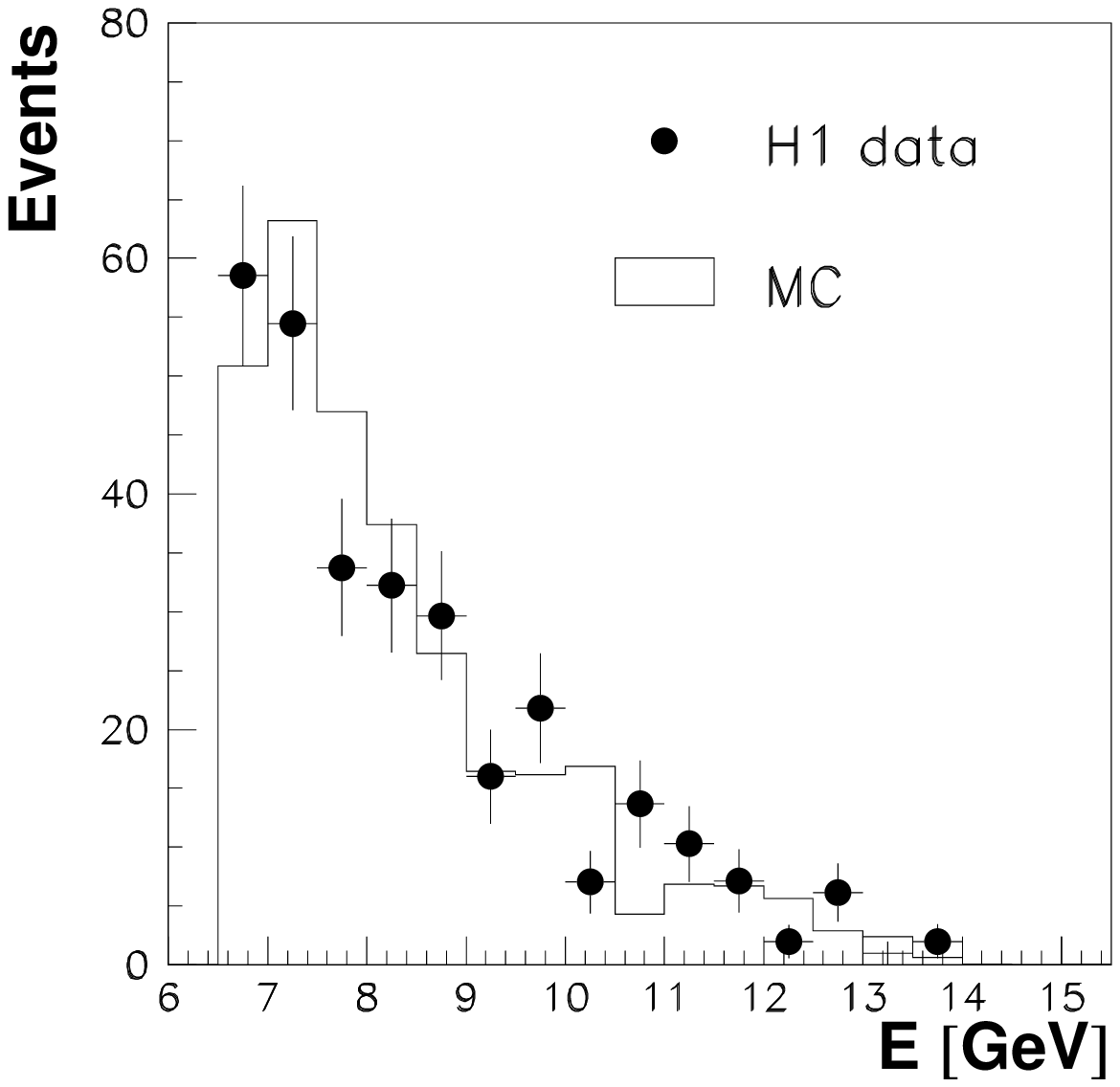,
width=10cm,height=15cm,bbllx=0pt,bblly=0pt,bburx=557pt,bbury=792pt}}
\put(30,-20){a)}
\put(60,-220){\epsfig{file=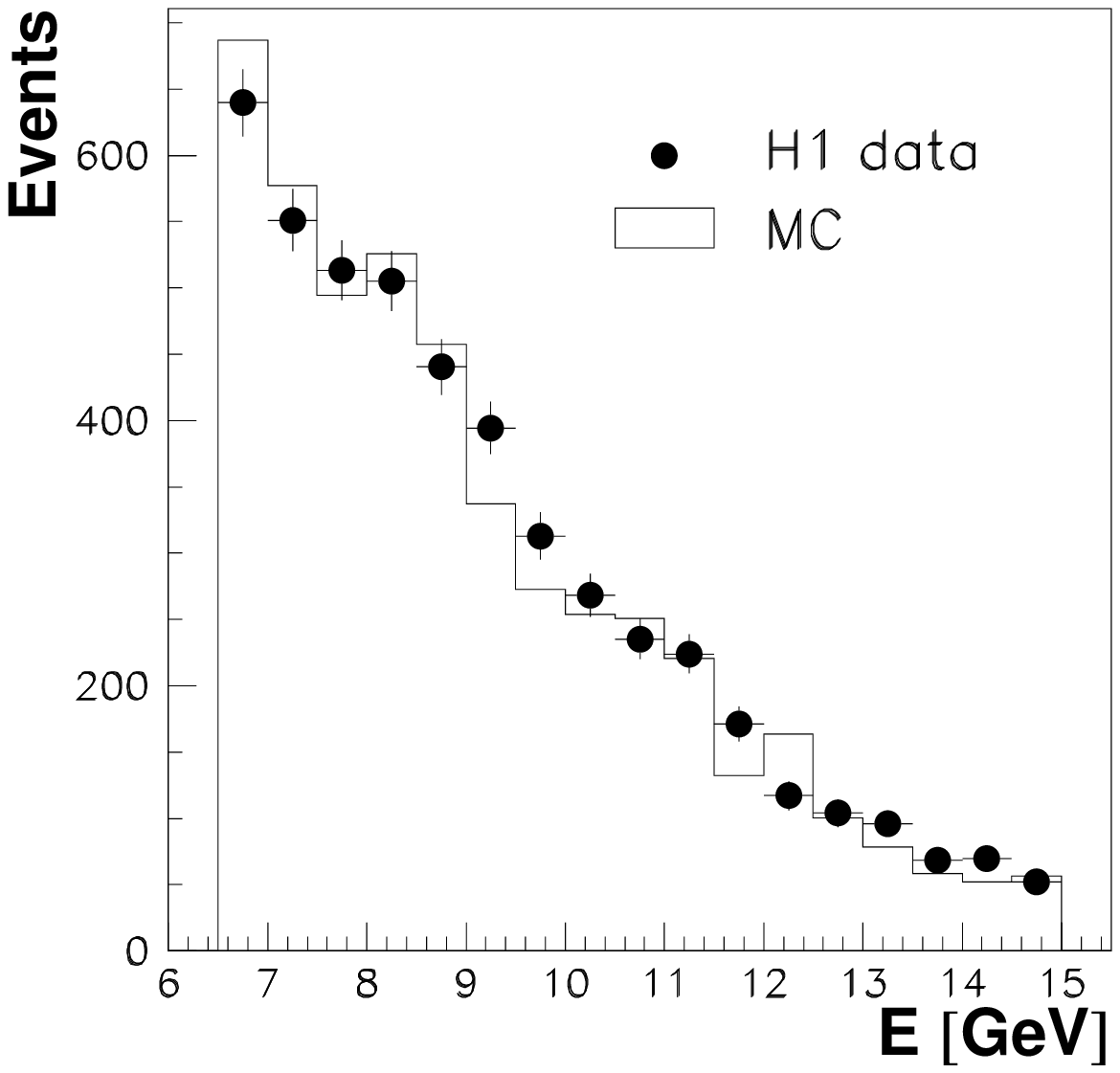,
width=10cm,height=15cm,bbllx=0pt,bblly=0pt,bburx=557pt,bbury=792pt}}
\put(260,-20){b)}
\end{picture}
\vspace*{4.0cm}
        \caption {\sl{ Energy distributions 
        of the highest energy cluster in the BEMC a)  
        for photoproduction events in which the scattered         
        positron was tagged and b) for
        the events rejected by the CIP requirement.
        The simulated spectra are  normalized 
        to the luminosity of the data.}}
        \protect\label{egp}
\end{figure}

The Monte Carlo simulation was used to 
subtract bin by bin the remaining photoproduction background.
A fraction of photoproduction interactions had the genuine 
final state positron detected  (``tagged'') in the electron tagger.
 Fig. \ref{egp}a shows the BEMC energy cluster
distribution for such events which passed the selection criteria.
Within the accepted range in energy (6.5 to 11 GeV) the simulation
reproduces the observed rate  to within 4\%.  

A further study was based on events with large
energy cluster radius ($\epsilon_1$)  or
 without CIP validation ($\epsilon_3$). These
samples predominantly consist of photoproduction events. The
event sample rejected by the $\epsilon_1$
cut (Table 1) allows the study of  faked positron signals by charged
hadrons. The shape of the energy spectrum
 agrees well with the simulated distribution and the
normalization is reproduced to within 7\%.
The event sample rejected by the $\epsilon_3$ 
cut allows the study of $\pi^0$
induced background. The  fake positron energy distribution of events in this 
sample is shown in Fig. \ref{egp}b. The normalization
 agrees to  within 2\%  of the
simulated rate. 

 The photoproduction background amounts to
$<20\% $ for the lowest $Q^2$ interval and decreases to 
$<5\% $ for the highest $Q^2$ interval. The normalization
uncertainty was estimated to be $20\%$
taking into account the fluctuations per bin of the
simulated sample and also the fact that only  
 part of the
photoproduction events,  with positron energies between 5~GeV
and 15~GeV, is tagged. 
\subsection{Cross Section Determination}
The deep inelastic scattering cross section was obtained 
by correcting the background subtracted 
number of events with  the  acceptance calculated from the Monte Carlo 
events, normalized to the measured luminosity. 
The cross section was corrected
for higher order QED and  electroweak contributions
 using the HECTOR \cite{HECTOR}
program. Starting from the GRV \cite{GRV} parton distributions, a two step
 iterative analysis 
was performed  to calculate the
acceptance and the radiative corrections. This maintains the
uncertainty of the cross section measurement due to input structure 
function variations below 1\%.

The radiative corrections were calculated  to order
$\alpha^2$ with soft photon exponentiation \cite{HECTOR, dima}. 
Taking into account the hadronic track requirements they are  about 35\%.
 Detailed
comparisons were made between the HECTOR result and the 
 HERACLES \cite{HERAKLES} Monte Carlo simulation which  
showed agreement at the per cent level. 
A study has also been made comparing the cross section
results with and without a selection $y_h > 0.1$ which, when
applied, reduces the
radiative corrections to about 15\%. The resulting cross sections
agreed to within 2\%.

The systematic error on the 
cross section is derived from the following contributions:
\begin{itemize}
\item{A 1\% uncertainty of the BEMC energy scale \cite{BEMC} leads to an error 
of about 1.5\%.}
\item{A 1~mrad uncertainty of the measured polar angle of the positron
causes a 2\% error.}
\item{The radiative corrections lead to a cross section uncertainty of 2\%.} 
\item{The vertex
 reconstruction efficiency is known to 2\% apart from the lowest $Q^2$ 
 interval where a 4\% error was estimated.}
\item{As in \cite{H1F} an efficiency error of 2\% is assigned to account
 for global event selection, BPC efficiency and TOF veto uncertainties.}
\item{The various cut efficiencies, studied using different deep inelastic
 and background enriched data and simulated samples, lead
 to an estimated systematic error of 3\% including the trigger
 efficiency error.}
\item{An extra error of 1\%  is estimated for positron 
misidentification  effects using the Monte Carlo simulation  and the
 study of the stability of the measurement against ignoring or considering
 the second highest energy cluster in the BEMC.}
\item{The photoproduction background was known  to
 within 20\%. This leads to a 4\% error at
 $Q^2=8.5~$GeV$^2$
 decreasing to 1\% at $Q^2=35$~GeV$^2$.}
\end{itemize}
Statistical errors in the Monte Carlo acceptance and efficiency 
calculations were computed and added quadratically to the systematic 
error.
The total systematic error on the cross section
 is about 8\% at $Q^2=8.5~$GeV$^2$ and
about 6\% for the higher $Q^2$ values. 
 Most of the error sources scale with $E_e'$ and are only
weakly dependent on $\theta_e$ in the range of the $F_L$
determination.
The statistical error  is about three times smaller
than the systematic error. 
\begin{table}[h]
\begin{center}
\begin{tabular}{|c|c||c|c|c||c|c|c|c|}
\hline
$   Q^2/$GeV$^2$ & $x$ & $\kappa \sigma$ &$ \Delta_{stat}$ &
$\Delta_{syst}$ & $F_2$ & $\Delta_{stat}$ & $\Delta_{syst}$ & $R_{calc}$ 
  \\    
\hline 
8.5   & 0.000135 & 1.165 & 0.027 & 0.095 & 1.354& 0.031 &0.110 &0.45    \\
12.0  & 0.000190 & 1.198 & 0.026 & 0.075 & 1.375& 0.030 &0.086 &0.40    \\
15.0  & 0.000238 & 1.368 & 0.032 & 0.079 & 1.561& 0.037 &0.090 &0.38    \\
20.0  & 0.000317 & 1.276 & 0.034 & 0.071 & 1.445& 0.038 &0.080 &0.35    \\
25.0  & 0.000396 & 1.439 & 0.042 & 0.079 & 1.651& 0.048 &0.091 &0.39    \\
35.0  & 0.000554 & 1.435 & 0.062 & 0.077 & 1.634& 0.071 &0.088 &0.37    \\
\hline
\end{tabular}
\end{center}
\caption{\label{CROS}\sl{
Inclusive cross section $\sigma=d^2 \sigma/dxdQ^2$,
 eq.(1), scaled by the kinematic 
factor $\kappa=Q^4x/(2 \pi \alpha^2 \cdot Y_+)$ with statistical
and systematic errors.
 The $Q^2,x$ values correspond to $y=0.7$ in  all bins. 
 Also quoted are the values of $F_2$ corresponding to these
cross section measurements
 with calculated $R$ values, given in the
rightmost column.
The values of $R=R_{calc}$ were obtained using the GRV parton
distributions [16] as input. 
There is  an additional, overall normalization uncertainty
of 1.5\% due to the luminosity measurement error.}}
\end{table}

\begin{figure}[b]\centering
 \begin{picture}(160,300)
\put(-170,350){
\epsfig{file=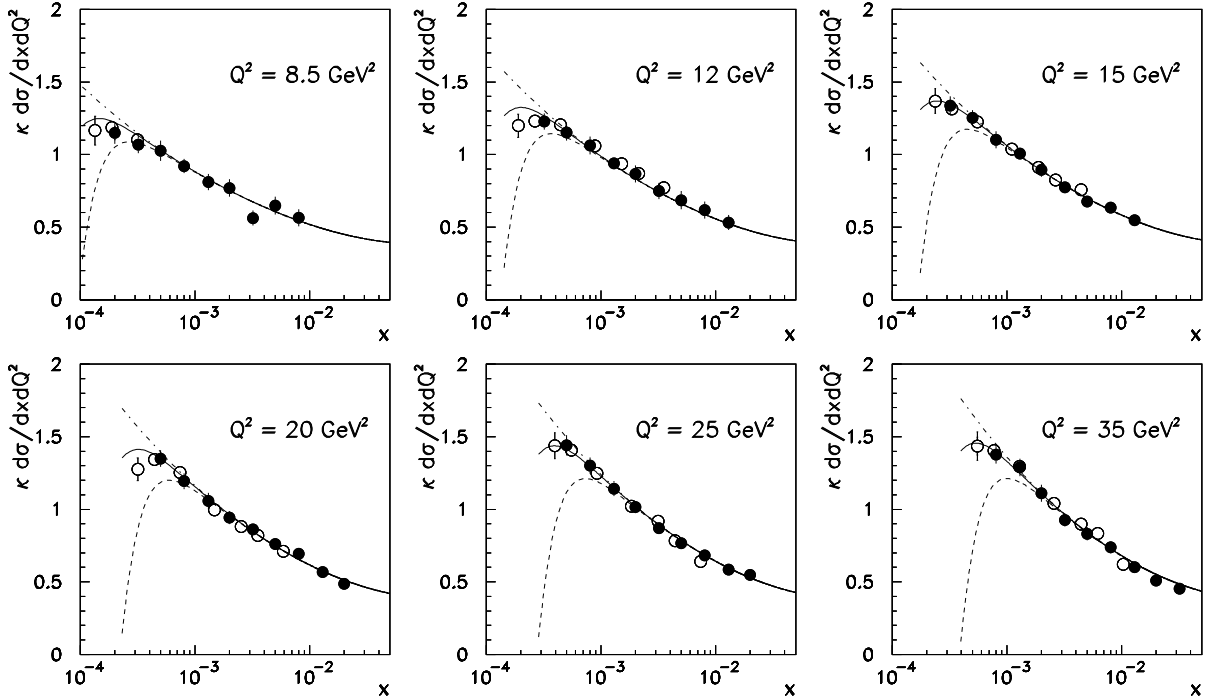,bbllx=0pt,bblly=0pt
,bburx=557pt,bbury=792pt,width=12cm,height=17cm,angle=-90.}}
\end{picture}
  \caption {\sl{Double differential cross section
    $ \kappa d\sigma /dxdQ^2   = F_2 -\frac{y^2}{Y_+} F_L$
    with $\kappa  = Q^4x/(2 \pi \alpha^2 \cdot Y_+)$ in 
    six $Q^2$ bins as a function of $x$.
    For $y > 0.6$ this analysis (open points) extends the
    previously published
    measurement [6] (closed points) towards lower $x$
    and is drawn here with full errors. The open
    points at larger $x$ are given without errors for ease of
    comparison with
    the data of [6]. The three lines represent calculated  
    cross sections  using for $F_2$ the QCD fit, 
    as described in sect. 3.1, and three different assumptions for 
    $F_L$. These are the two extremes,
    $F_L=0$ (dashed-dotted line) and $F_L=F_2$ using 
    $F_2$ from the QCD fit (dashed
    line), and
    $F_L$ as calculated in NLO from the quark and gluon distributions
    determined by the QCD fit (solid line).}}
        \protect\label{sigm}
\end{figure}

The $ep$ cross section  is given in Table \ref{CROS}
for the six new intervals in $Q^2$ and $x$. 
The  analysis was also  extended into the region of our
previously published $F_2$
results for $0.6 > y > 0.03$. In  Fig. \ref{sigm} the present cross section 
measurement is
shown together with the data   
\cite{H1F}. There is everywhere good
agreement in the region of overlap.
The cross section is quoted in the form
 $\sigma \cdot Q^4 x /(2 \pi \alpha^2 \cdot Y_+) = F_2 - y^2 F_L/Y_+$ 
which for small $y$ is about equal to $F_2$ independently of $F_L$.
The three lines drawn in Fig. 3 represent cross sections calculated
using the QCD fit for $F_2$ which is described in the subsequent section
and three different assumptions on $F_L$. The dashed-dotted and
dashed  lines correspond to the limits $F_L=0$ and $F_L=F_2$,
respectively,  as required
 by the positivity of the cross sections  $\sigma_L$ and $\sigma_T$. 
The solid line represents the cross section with $F_L$ calculated 
using the gluon and quark distributions obtained by
the QCD analysis of $F_2$. It becomes apparent in
Fig. 3 that at lowest $x$, corresponding to the high $y$ region of
this data, the cross section becomes very sensitive to the
longitudinal structure function. On the contrary, at larger $x$, for
about $y < 0.35$, the $F_2$ contribution dominates and the
three lines nearly coincide. Most of the previously
published $F_2$ data points
are insensitive to the assumptions on $F_L$. 

In the publication \cite{H1F} the measured DIS cross section was
used to determine the structure function $F_2$ assuming $R$ to
be given by the GRV
parton distributions~\cite{GRV}
using the relation \cite{am}. This measurement represents an
extension of the previous cross section data towards lower $x$.
In order to provide a consistent set of structure function values
the corresponding six
values of $F_2$ were derived following the same procedure
 (see  Table 2). Note that these values are  rather
sensitive to the $R$ values chosen.
\section{Determination of $F_L$}
\subsection{QCD Fit}
For the subtraction of the $F_2$ contribution to the cross section a 
NLO QCD fit was performed using the DGLAP evolution 
equations. The fit used  the  H1 data
\cite{H1F}  for $y < 0.35$.
The BCDMS proton and deuterium data  \cite{BCDMS} were used to constrain the
high $x$ behaviour of the parton distributions. In contrast with the
previous QCD analysis performed by H1 \cite{H1F}, the NMC data
\cite{NMC} were not
included in the standard fit  to ensure a maximum 
weight of the H1 data in the fit procedure. The starting point of the 
evolution was chosen to be $Q^2_0=5$~GeV$^2$ and all data with $Q^2 \geq
 Q^2_{min}=1.5$~GeV$^2$ were included in the fit. 
To avoid possible  higher twist effects, BCDMS data in the range  
 $ x > 0.5$ for $Q^2<15~$GeV$^2$
were not included in the fit. The normalization of the H1 data was
kept fixed.
 The fit used three light flavors with the charm
contribution added using the NLO calculation of the photon-gluon fusion
process \cite{GRV, riem}. Furthermore, the momentum sum rule 
was imposed and the integral over the valence quark
distributions was set to 2 for $u_v$ and to 1 for $d_v$. The input parton
 distributions  at the starting scale $Q^{2}_{0}$ were parameterized as follows:
\begin{eqnarray}\label{input}
xg(x)&=& A_gx^{B_g}(1-x)^{C_g},\nonumber \\
xu_v(x)&=& A_{u}x^{B_{u}}(1-x)^{C_{u}}(1+D_{u}x+E_{u}\sqrt{x}),\nonumber \\
xd_v(x)&=& A_{d} x^{B_{d} }(1-x)^{C_{d}}(1+D_{d}x+E_{d}\sqrt{x}),\nonumber \\
xS(x)&=& A_{S} x^{B_{S} }(1-x)^{C_{S}}(1+D_{S}x+E_{S}\sqrt{x}),
\end{eqnarray}
where  $S=  \bar{u} =  \bar{d} = 2 \bar{s}$ defines the sea distributions.
Three fits with different, fixed $\Lambda_{QCD}$ values were performed.
The best  $\chi^2 /ndf$ of 506/(505-15) was obtained 
for $\Lambda_{QCD}=210~$MeV.   The fitted parton distribution 
functions were evolved into the new domain using the NLO DGLAP 
equations and used to calculate the corresponding values of $F_{2}$.

Table \ref{fitu} summarizes the $Q^2$ averaged uncertainties 
in $F_{2}$ arising from the fit procedure. 
\begin{table}[h]
\begin{center}
\begin{tabular}{|l|r|}
\hline
 fit assumption & uncertainty in \%                     \\   
\hline 
NMC data used                         &  1.4       \\
change of $\Lambda_{QCD}$ by 50~MeV   &  0.7       \\
$g(x,Q^2_o) \cdot (1+E \sqrt{x})$     &  0.1       \\
$Q^2_{min}=5$~GeV$^2$                 &  0.6       \\
$Q^2_o = 3~$GeV$^2$                   &  0.4       \\
\hline
\end{tabular}
\end{center}
\caption{\label{fitu}\sl{
Uncertainty, relative to the result of the standard fit (see text),
 of the  structure function $F_2$ 
averaged over the $Q^2$ range of the $F_L$ data for various 
assumptions in the QCD fit procedure.}}
\end{table}
The total uncertainty due to the fit assumptions
amounts to 1.7\%. The resulting absolute error
of the longitudinal structure function
$\Delta F_L \simeq  Y_+/y^2 \cdot \Delta F_2$ is 
approximately 0.07. 

There is  a  small dependence of the
structure function $F_{2}$ on $F_L$ due to the 
assumption  made for $F_L$ in the cross section analysis for $y < 
0.35$. Thus the two extreme assumptions $F_{L}=0$ and
$F_{L}=F_{2}$ were used and two modified structure functions $F_{2}$ 
were derived as input to  two QCD fits.
This changed the QCD predicted $F_2$ at $y=0.7$ on average by  $-1.6$\% 
and $+3.8$\%, respectively. Thus  an asymmetric error on $F_L$
was introduced and
 was added in quadrature to the other systematic errors. 

Two different cross checks were made of the prediction of $F_2$
at lowest $x$ by using the perturbative dipole model with $k_T$
factorization \cite{dipole} and an empirical model
based on the similarity of the rise of $F_2$ at low $x$ and
 the evolution of the charged multiplicity
with energy in $e^+e^-$ collisions \cite{monopol}.
The model parameters were determined using the previously
published
H1 $F_2$ data \cite{H1F} for $y<0.35$ and $Q^2$ boundaries given
by the limitations of these approaches.  The three-parameter
 $F_2$ function in the dipole model,
 calculated at $y=0.7$,  is only  2\% lower
than the structure function
obtained by the evolution procedure described above. 
Similarly, the two-parameter $F_2$ function of the
empirical model  is on average  2\%
higher than the QCD fit result at $y=0.7$. 
Thus both approaches to extrapolate $F_2$ would lead to a result for the
longitudinal structure function in very good agreement with 
the one obtained subsequently using the QCD fit for the description of $F_2$.

\subsection{Results}
The measured longitudinal structure function $F_L$ is given in Table 
\ref{tfl}. The systematic error consists of the following contributions:
\begin{itemize}
 \item{The experimental errors of the cross section measurement which are
  uncorrelated with the error of the data entering the QCD fit at
  lower $y$. These are error sources, such as the tracking trigger
  or CIP efficiency, which are specific for the high $y$ range   and
  to this analysis.}
 \item{The  error  due to  possible variations of the assumptions in 
  the QCD fit procedure as discussed in Section 3.1.}
 \item{The experimental errors like energy and angle uncertainties and
  global efficiency and luminosity errors which are mostly common to both
  the low and the high $y$ region. A correlation of these errors is
  introduced through the QCD fit of $F_2$. This error source includes
  also the statistical error of the fit result.}
\end{itemize}

  The third contribution comprises several effects. For example, 
  any global shift common to the high $y$ data and the H1 data used in the fit,
  like the luminosity uncertainty, gets reduced to about 1/3 of its 
  magnitude.  A reduction of the error is observed as well for the 
  polar angle uncertainty. However, the error of the energy of the
  scattered  positron is not compensated. 
  The  H1 $F_2$ data \cite{H1F} for  $y \geq 0.15$ were
  obtained with the electron method and those at  smaller 
  $y$  with the sigma method. A 1\% increase in $E_e'$ increases the cross
  section for $y \geq 0.15$  and decreases it below that value,
  a behaviour which leads to a large change in the $\chi^2$ of the
  fit. Thus the   measurement of the positron energy  
 is the dominating ``correlated" error although it has only a small
 effect on the cross section at high $y$. 
\begin{table}[h]
\begin{center}
\begin{tabular}{|c|c||c|c|c|c||c|c|c|}
\hline
$   Q^2/$GeV$^2$ & $x$ & $F_L$ &$ \Delta_{stat}$ & $+\Delta_{syst}$ &
 $ -\Delta_{syst}$ 
 & $\Delta_{exp}^{unc} $ & $\Delta_{exp}^{cor}$ & $ \Delta_{fit}$  \\    
\hline 
8.5   & 0.000135 & 0.51 & 0.06 & 0.29 & 0.27 & 0.17 & 0.19 & 0.06   \\
12.0  & 0.000190 & 0.63 & 0.06 & 0.28 & 0.25 & 0.15 & 0.18 & 0.07   \\
15.0  & 0.000238 & 0.35 & 0.08 & 0.29 & 0.27 & 0.15 & 0.19 & 0.08   \\
20.0  & 0.000317 & 0.67 & 0.08 & 0.28 & 0.26 & 0.14 & 0.18 & 0.07   \\
25.0  & 0.000396 & 0.33 & 0.10 & 0.25 & 0.22 & 0.15 & 0.14 & 0.07   \\
35.0  & 0.000554 & 0.39 & 0.15 & 0.24 & 0.21 & 0.14 & 0.14 & 0.07   \\
\hline
\end{tabular}
\end{center}
\caption{\label{tfl}\sl{
The longitudinal structure function $F_L$ 
with statistical ($\Delta_{stat}$) and systematic errors
($\pm \Delta_{syst})$:
$\Delta_{exp}^{unc}$ is the uncorrelated experimental
cross section error at high $y$,
$\Delta_{exp}^{cor}$ is  the correlated experimental 
error 
 and $\Delta_{fit}$ is the error introduced by the 
 QCD fit uncertainty. The total systematic error contains also
 the asymmetric contribution due to the assumptions on $F_L$
 in the determination of $F_2$, see Section 3.1.
 The $Q^2,~x$ values correspond to $y=0.7$ in all bins.}}
\end{table}

\begin{figure}[ht]\centering
\begin{picture}(160,95)
\put(-150,-270){
\epsfig{file=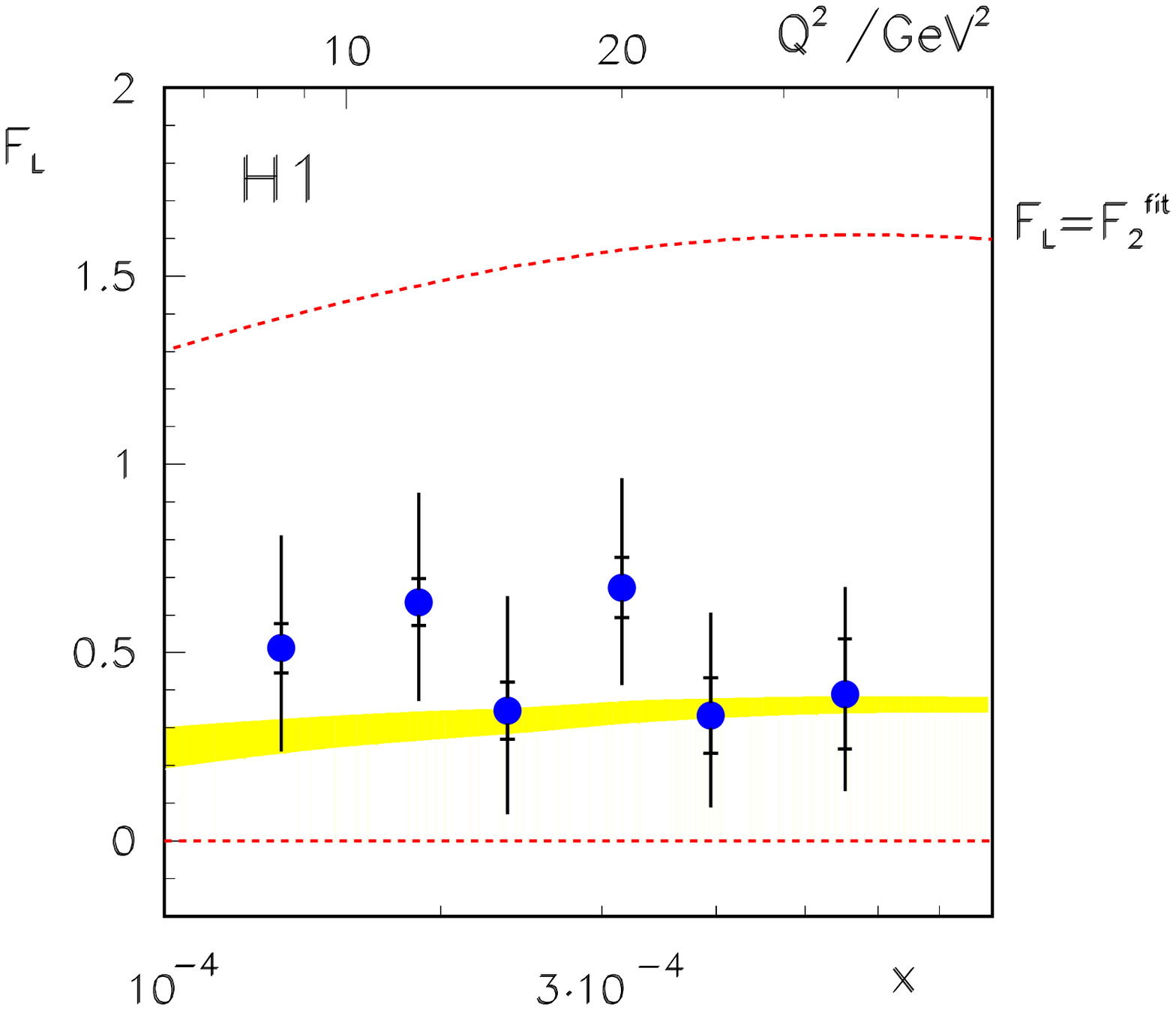,
width=15cm,height=18.5cm,bbllx=0pt,bblly=0pt,bburx=557pt,bbury=792pt}}
\end{picture}
\vspace*{6cm}
      \caption {\sl{Longitudinal structure function
     $F_L$  as 
    function of $Q^2$ or $x=Q^2/sy$
    for $y=0.7$. The inner error bars are the
     statistical 
    errors. The  full error bars represent the statistical and
    systematic errors added in quadrature.
    The error band represents the uncertainty of the 
    calculation of $F_L$ using the gluon and quark distributions,
    as determined from the NLO QCD analysis 
     of the H1 data [6] for $y \leq 0.35$ 
     and the BCDMS data [8].
     The dashed lines define the allowed range of $F_L$ values from
     $F_L=0$ to $F_L=F_2$ where $F_2$ is given by the QCD fit.}}
        \protect\label{fl}
\end{figure}

The six measurement values enable a determination to be made of the mean
$F_L$ and its derivative $d F_L/ d \ln (x)$
for $Q^2= 0.7 \cdot sx$ from a straight line $F_L = a + b \cdot \ln (x)$. 
Taking into
account the error correlations between the six data points we obtain
a mean  $F_L = 0.52
  \pm 0.03$~(stat)$^{~+0.25}_{~-0.22}$ (syst) and a derivative
  $ d F_L/ d \ln (x) = -0.085
  \pm 0.080 $~(stat)$^{~+0.082}_{~-0.083}$ (syst)
at  $Q^2=15.4~$GeV$^2$ and $x=0.000243$. Note that
the derivative  has comparable statistical and systematic errors
while the error of the mean $F_L$ is  dominated by systematics.
Fig.~\ref{fl} shows the data of Table \ref{tfl} and the extreme limits
of $F_L=0$ and $F_L=F_2$ using the QCD fit.
Without
utilizing the measured dependence on $x$ or $Q^2$,
these extremes are excluded with 2.3 and
4.0 times the total error, respectively.
%
%

At low $x$ the longitudinal structure function is related to the gluon
distribution.  The dashed band
in Fig.~4 represents the calculation of $F_L$ according to  
\cite{vN} for  three light quarks 
and according to \cite{riem} for the charm contribution.
The input gluon and quark 
distributions are determined by the NLO QCD fit described in Section
3.1. The width of this band is determined 
by the experimental errors of the $F_2$ data, taking into account
their point-to-point correlations, and by the
fit uncertainties discussed above.
At the present level of accuracy
there is  consistency between the structure function $F_L$
determined from  this analysis
and that calculated from  the gluon and quark distributions.
\section{Summary}
Based on data taken in 1994 with a luminosity of 1.25~pb$^{-1}$, an 
inclusive measurement of the deep inelastic cross section measurement
at $y=0.7$  has been used to 
determine for the first time the longitudinal structure function 
$F_L(x,Q^2)$ at very low Bjorken $x$. The analysis assumed the proton 
structure function $F_2(x,Q^2)$ to be in accordance with 
next-to-leading order perturbative 
QCD. The result excludes the
extreme limits of $F_L=0$ and $F_L=F_2$, corresponding to $R=0$ and 
$R=\infty$, by 2.3 and 4.0 times the total error on $F_L$.
 The result is consistent with a higher order
 QCD calculation of $F_L$ which 
essentially relied on the gluon distribution as determined from
the $F_2$ structure function data. 
 \\  \\
{\bf Acknowledgments}\\ \\
 We are very grateful to the HERA machine group whose outstanding
 efforts made this experiment possible. We acknowledge the support
 of the DESY technical staff.
 We appreciate the big effort of the engineers and
 technicians who constructed and maintain the detector. We thank the
 funding agencies for financial support of this experiment.
 We wish to thank the DESY
 directorate for the support and hospitality extended to the
 non-DESY members of the collaboration.  Finally, helpful discussions
 with D.Yu. Bardin are acknowledged.     
¥
\newpage

\begin{thebibliography}{99}
        \bibitem{CG}
        C.G. Callan and D. Gross, Phys. Rev. Lett. {\bf 22} (1969) 156.

        \bibitem{AM}
        A. Zee, F. Wilczek and S.B. Treiman, Phys. Rev. {\bf D 10}
        (1974) 2881.

        \bibitem{vN}
        E.B. Zijlstra and W. van Neerven, Nucl. Phys. {\bf B 383}
        (1992) 525; \\
        S.A. Larin and J.A.M. Vermaseren, Z.Phys. {\bf C 57} (1993) 93.

        \bibitem{Bo}
        L. Whitlow et al., Phys. Lett. {\bf B 250} (1990) 193; \\
        A.Bodek,
          Proceedings of the 4th International
          Conference on Deep Inelastic Scattering, Rome, April 1996,
          to be published, and references cited therein.

        \bibitem{Mil}
        NMC Collaboration, M.~Arneodo et al., subm. to
        Nucl. Phys.~{\bf B}, hep-ph/9610231.

        \bibitem{H1F}
          H1 Collaboration, S.~Aid et al., Nucl.Phys.~{\bf B 470} (1996) 3.       

        \bibitem{DGLAP}
         Yu.L. Dokshitzer, Sov. Phys. JETP {\bf 46} (1977) 641; \\
         V.N. Gribov and L.N. Lipatov, Sov. J. Nucl. Phys.
        {\bf 15} (1972) 438 and 675; \\
         G. Altarelli and G. Parisi, Nucl. Phys. {\bf B 126} (1977) 298.

     \bibitem{BCDMS} 
      BCDMS Collaboration, A.C. Benvenuti et al., Phys. Lett. {\bf B 223}
      (1989) 485; \\
              CERN preprint CERN-EP/89-06 (1989).
     
        \bibitem{H1}
        H1 Collaboration, S. Aid et al., DESY H1 note 96-01(1996),
        Nucl. Instr. and Meth. {\bf A}, to appear.

       \bibitem{django}
       G.A. Schuler and H. Spiesberger, Proceedings of the Workshop
       Physics at HERA, DESY 1992, Eds. W. Buchm\"uller and G. Ingelman,
       vol.3, 1419.

        \bibitem{HERAKLES}
        A. Kwiatkowski, H. Spiesberger and H.-J. M\"ohring,
        Comput. Phys. Comm. {\bf 69} (1992) 155.

        \bibitem{LEPTO}
        G. Ingelman,  Proceedings of the Workshop
       Physics at HERA, vol. 3, Eds. W. Buchm\"uller and G. Ingelman,
       DESY (1992) 1366.  

       \bibitem{PHOJET}
       R. Engel, Proceedings of the XXIXth Rencontres de Moriond,
       Ed. J. Tran Thanh Van (Editions Fronti\`eres, 1994) 321.

       \bibitem{GEANT}
        R. Brun et al., GEANT3 User's Guide, CERN-DD/EE 84-1, Geneva (1987).
           
        \bibitem{HECTOR}
        A.~Arbuzov et al., Comput. Phys. Comm. {\bf 94} (1996) 128.

        \bibitem{GRV} 
        M. Gl\"uck, E. Reya and A. Vogt, Z. Phys. {\bf C 67} (1995) 433.

        \bibitem{dima}
        D.Yu. Bardin et al., Proceedings of the  Workshop on HERA
        Physics, DESY 1996, ed. by G. Ingelman, A. De Roeck
        and R. Klanner, vol.1, 13,
        DESY 96-198,  hep-ph/9609399.

       \bibitem{BEMC}
        H1 BEMC Group, J. B\'{a}n et al.,
        Nucl. Instr. and Meth. {\bf A 372} (1996) 399.

        \bibitem{am}
       G.~Altarelli and G.~Martinelli, Phys. Lett.~{\bf B 76} (1978) 89.
       
       \bibitem{NMC}
        NMC Collaboration, M. Arneodo et al., Phys. Lett. {\bf  B 364}
       (1995) 107.

       \bibitem{riem}
       E. Laenen et al., Nucl.\ Phys.\ {\bf B 392} (1993) 162.
   
       \bibitem{dipole}
       A. Mueller, Nucl. Phys. {\bf B 415} (1994) 373;  \\
       H. Navelet et al., Phys. Lett. {\bf B385} (1996) 357.

       \bibitem{monopol}
       A. De Roeck and E.A. De Wolf, DESY preprint 96-143 (1996),
       hep-ph/9609203.    
  

 \end{thebibliography}
 \end{document}